\documentclass[twocolumn,english,aps,prl,reprint,nofootinbib,superscriptaddress]{revtex4-2}
\pdfoutput=1
\usepackage[T1]{fontenc}
\usepackage[latin9]{inputenc}
\usepackage{amsmath}
\usepackage{amssymb}
\usepackage{graphicx}

\usepackage{color}
\usepackage{float}
\usepackage{hyperref}

\hypersetup{
    colorlinks,
    linkcolor={black},
    citecolor={black},
    urlcolor={blue},
}

\makeatletter
\@ifundefined{textcolor}{}
{%
 \definecolor{BLACK}{gray}{0}
 \definecolor{WHITE}{gray}{1}
 \definecolor{RED}{rgb}{1,0,0}
 \definecolor{GREEN}{rgb}{0,1,0}
 \definecolor{BLUE}{rgb}{0,0,1}
 \definecolor{CYAN}{cmyk}{1,0,0,0}
 \definecolor{MAGENTA}{cmyk}{0,1,0,0}
 \definecolor{YELLOW}{cmyk}{0,0,1,0}
 \definecolor{PURPLE}{rgb}{0.7,0,0.7}
 \definecolor{dgreen}{rgb}{0,0.6,0}
}

\newcommand{\blue}{\textcolor{black}}

\usepackage{pslatex}

\usepackage{babel}

\makeatother

\begin{document}

\title{Fast and High-Yield Loading of a D$_2$ MOT of Potassium from a Cryogenic Buffer Gas Beam}

\author{Zack Lasner}
\email{zlasner@g.harvard.edu}
\affiliation{Harvard-MIT Center for Ultracold Atoms, Cambridge, MA 02138, USA}
\affiliation{Department of Physics, Harvard University, Cambridge, MA 02138, USA}

\author{Debayan Mitra}
\affiliation{Harvard-MIT Center for Ultracold Atoms, Cambridge, MA 02138, USA}
\affiliation{Department of Physics, Harvard University, Cambridge, MA 02138, USA}

\author{Maryam Hiradfar}
\affiliation{Harvard-MIT Center for Ultracold Atoms, Cambridge, MA 02138, USA}
\affiliation{Department of Physics, Harvard University, Cambridge, MA 02138, USA}

\author{Benjamin Augenbraun}
\affiliation{Harvard-MIT Center for Ultracold Atoms, Cambridge, MA 02138, USA}
\affiliation{Department of Physics, Harvard University, Cambridge, MA 02138, USA}

\author{Lawrence Cheuk}
\altaffiliation{Present address: Department of Physics, Princeton University, Princeton, NJ 08544 US}
\affiliation{Harvard-MIT Center for Ultracold Atoms, Cambridge, MA 02138, USA}
\affiliation{Department of Physics, Harvard University, Cambridge, MA 02138, USA}

\author{Eunice Lee}
\altaffiliation{Present address: Department of Physics, Massachusetts Institute of Technology, Cambridge, MA 02139, USA}
\affiliation{Harvard-MIT Center for Ultracold Atoms, Cambridge, MA 02138, USA}
\affiliation{Department of Physics, Harvard University, Cambridge, MA 02138, USA}

\author{Sridhar Prabhu}
\altaffiliation{Present address: Laboratory of Atomic and Solid State Physics, Cornell University, Ithaca, New York, USA}
\affiliation{Harvard-MIT Center for Ultracold Atoms, Cambridge, MA 02138, USA}
\affiliation{Department of Physics, Harvard University, Cambridge, MA 02138, USA}

\author{John Doyle}
\email{jdoyle@g.harvard.edu}
\affiliation{Harvard-MIT Center for Ultracold Atoms, Cambridge, MA 02138, USA}
\affiliation{Department of Physics, Harvard University, Cambridge, MA 02138, USA}

\date{\today}

\begin{abstract}
We demonstrate the direct loading of a D$_2$ MOT of potassium-39 atoms from a cryogenic buffer gas beam source. We load \blue{$10^8$}~atoms in a 10~ms pulse, with no degradation in performance up to a 10~Hz repetition rate. Observed densities reach \blue{$\sim10^{11}$}~atoms/cm$^3$ in a single pulse, achieved with no sub-Doppler cooling or transverse compression. This system produces an ideal starting point for ultracold atom experiments where high experimental repetition rates are desirable and initial high densities are critical. Extension to other atomic species (e.g., refractory metals) that present technical challenges to high-yield production using oven-based sources is straightforward.

\end{abstract}
\maketitle

Ultracold atomic systems have been at the core of developments in the fields of quantum simulation~\cite{bloch2008many,Bloch2012,Gross2017,Ueda2014,Goldman2014,Windpassinger2013,Stamper-Kurn2013,Lewenstein2007}, quantum information and computation~\cite{Georgescu2014,Vogel2011,Endres2016,Alexeev2019,Monroe2021,Bernien2017,Morgado2021}, frequency metrology~\cite{Zhang2016,Ludlow2015,Boyd2007,Ushijima2015,Nicholson2015,Pustelny2013,Bodine2020,Beloy2021,Derevianko2014,Derevianko2018,Huntemann2014}, and precision measurement and searches for physics beyond the Standard Model~\cite{Safronova2018,Cronin2009,Bassi2013,Geraci2016,Graham2016,Biedermann2015,Yu2019,Bennett2008,Bishof2016,Hamilton2015,Baker2021}. 
The first step in producing ultracold atoms usually begins with a hot vapor source at a few hundred kelvin. The thermal velocities of atoms in this vapor are distributed in a Maxwell-Boltzmann distribution with mean value usually $\sim$1000 m/s. To produce a magneto-optical trap (MOT), atoms are slowed down to velocities within the MOT capture range, typically $\sim$50-100 m/s. Zeeman slowing~\cite{Phillips1982Laser} achieves this, but the loading time-scale for a typical atomic MOT ranges between 1-20 seconds, limited by the flux of slow atoms at the trap site, the slowing length, and the background gas pressure. In applications where evaporative cooling~\cite{Ketterle1996} is required, the experimental preparation time after MOT loading is usually several seconds even with advanced modern techniques~\cite{Wilkowski2010,Ravensbergen2018,Kim2019,Hung2008,Otoishi2020,Ulitzsch2017} due partly to finite atomic density. Reducing the MOT loading time, while still achieving the needed number density, could provide a significantly improved experimental repetition rate. In many applications, where the ultracold temperatures and densities realized in a MOT are sufficient, the experimental repetition rate could be increased by at least an order of magnitude if the MOT loading time were reduced to $\sim$10-100 ms.

One approach to improving MOT loading is to sidestep ovens and use cold sources where the forward velocity and brightness are independent of the atom's boiling point. Cryogenic buffer gas beam (CBGB) sources are versatile sources of cold atoms and molecules that achieve extremely high instantaneous flux, while also having low forward velocity independent of chemical reactivity. 
Radical species produced in CBGBs to date include a wide variety of atoms~\cite{Maxwell2005,Hemmerling2014,Patterson2009}, diatomic molecules~\cite{hemmerling2016laser,truppe2017intense,barry2011bright,Lu2011,Albrecht2020,Skoff2011,Ding2020},  and polyatomic molecules~\cite{augenbraun2020laser,Baum2020,Mitra2020}. A similarly diverse array of non-radical species including atoms~\cite{Patterson2007,Hemmerling2014}, diatomic molecules~\cite{Maxwell2005,Hutzler2011,Patterson2007,petricka2007}, and polyatomic molecules~\cite{VanBuuren2009,Herschbach2009,Patterson2009,Sawyer2011,Eibenberger2017,Spaun2016,Satterthwaite2019,Porterfield2019,Yeh2019,Patterson2013a,Piskorski2014} has also been produced and studied. In the past, a cryogenic buffer gas beam source has been used to load MOTs of lanthanide atoms to demonstrate the feasibility of trapping atoms with imperfect cycling transitions \cite{Hemmerling2014}.


The near-universality of CBGBs makes them especially useful for producing ultracold refractory atoms, which have low vapor pressures and thus require very high oven temperatures. 
Laser cooling pathways have been identified in refractory species such as chromium~\cite{Griesmaier2005,Chicireanu2006,Bradley2000,Gabardos2019,Bismut2011}, titanium, zirconium, vanadium, niobium, manganese, technetium, and ruthenium, in addition to other atoms with low vapor pressures~\cite{Eustice2020}. An active experimental effort is underway toward a MOT of titanium~\cite{Neely2021}.

In this work, we demonstrate and study rapid loading of an atomic MOT at high density using $^{39}$K atoms as a test species. We demonstrate a high-flux, low-temperature pulsed atomic source, and pulsed loading of a three-dimensional MOT in a D$_2$ configuration in 10~ms at repetition rates up to 10~Hz.

\begin{figure*}[ht]
\begin{centering}
\includegraphics[width = 1\textwidth]{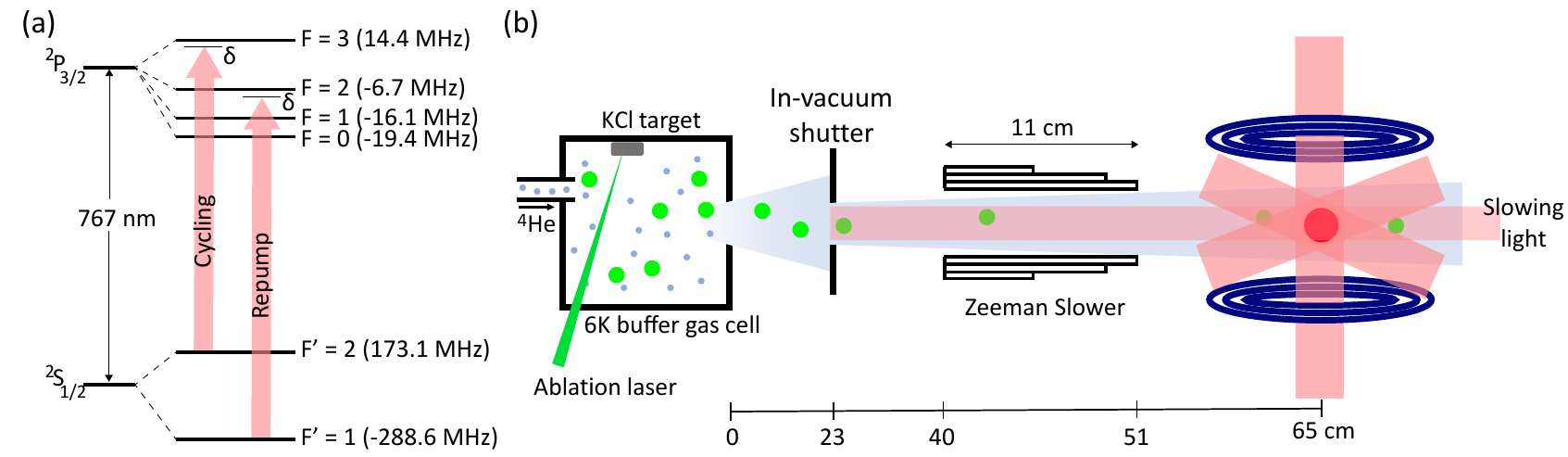}
\caption{\textbf{Experimental schematic and relevant laser frequencies.} (a) Potassium energy level diagram depicting the D$_2$ transition of $^{39}$K employed in this paper. Both the cycling and the repumping transitions were equally detuned from resonance by $\delta$ = \blue{-14}~MHz. (b) Schematic of the experiment depicting the CBGB source, the Zeeman slower, and the MOT (not to scale). The in-vacuum shutter is placed on the room temperature plate of the beam source vacuum chamber.}\label{fig:apparatus}
\par\end{centering}
\end{figure*}

The CBGB source is standard, like those described in detail elsewhere \cite{hutzler2012buffer,barry2011bright}. Briefly, a copper cell is thermally anchored to a pulse-tube cryo-cooler at a temperature of approximately 6 K. The cell has a cylindrical bore of 2.5 cm in diameter and 5 cm in length. Between 2 and 3 standard cubic centimeters per minute (sccm) of helium, also pre-cooled to 6 K, enters at the back of the cell and exits through a 7 mm diameter aperture in the front. A solid ``target'' of precursor material, in this case KCl, is ablated by the harmonic of a pulsed Nd:YAG laser at 532 nm, with up to 20 mJ per pulse. Potassium atoms released in the ablation plume are cooled by the helium buffer gas and hydrodynamically extracted into a beam out of the front cell aperture in a $1/e$ time of around 0.5~ms.

The KCl target is prepared by compressing KCl powder (Alfa Aesar, 99.995\% purity) with a hydraulic press at a pressure of $1.4\times10^5$ kPa. Since KCl crystals are translucent, we add a few drops of Zirconium powder suspended in water before compression to increase the opacity of the target and enhance energy deposition by the ablation laser. The potassium content is of natural isotopic abundance, hence greater than 93\% is $^{39}$K.

The beam line is depicted schematically in Fig.~\ref{fig:apparatus}(b). The room-temperature front face of the CBGB source chamber is equipped with an in-vacuum shutter (Uniblitz VS14E1T0-ECE) that limits the flow of buffer gas into the MOT chamber. Farther downstream is a short Zeeman slower, followed by the MOT chamber. The center of the MOT chamber is 65 cm from the cell exit. The cryogenic beam source is naturally cryo-pumped, and the chamber's exit plate is also pumped by a turbo pump with a pumping speed for He of 60~L/s. The beam line is equipped with a turbo pump with a pumping speed of 230~L/s,
an ion pump with a pumping speed of 50~L/s, and a titanium sublimation pump. The base pressure reached in the beam line prior to flowing any buffer gas is $\sim$$2\times10^{-9}$~Torr but that pressure increases to $\sim$$1\times10^{-7}$~Torr when flowing buffer gas. While that is still well below the level required for this work, it could be easily reduced further with simple differential pumping.

\begin{figure*}[ht]
\begin{centering}
\includegraphics[width = 1.5\columnwidth]{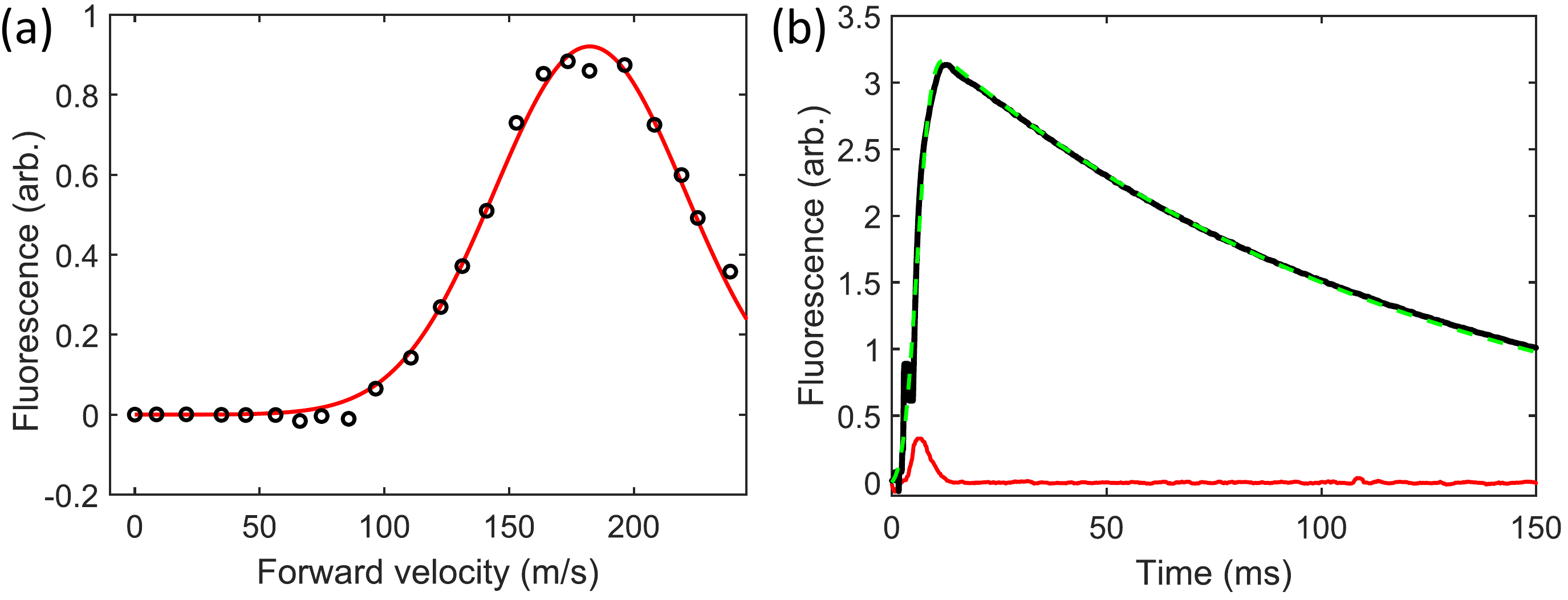}
\caption{\label{fig:beamtraces} \textbf{Forward velocity and MOT fluorescence}. (a) Potassium beam forward velocity measured at the MOT location. Peak velocity from a Gaussian fit to the data is 182~m/s, with negligible population at and below the MOT capture velocity of \blue{67}~m/s. Velocities beyond 250~m/s are not shown due to spurious signal in the velocity profile arising from the ground-state hyperfine structure. (b) PMT trace of an unperturbed atomic beam arriving at the MOT location is shown in red and the MOT trace is shown in black. The MOT trace is fit (dashed green line) to the theoretical profile given in Eq.~\ref{eq:profile}. We have fully loaded the MOT by 12~ms after ablation and measure a MOT lifetime of 116.4(4)~ms.}
\par\end{centering}
\end{figure*}

The laser system consists of two tapered amplifier systems (Toptica TA Pro) operating at 767~nm. 
We employ a potassium vapor cell and perform modulation transfer Doppler-free spectroscopy to lock the first laser to the crossover feature between the two hyperfine sublevels of the $^2S_{1/2}$ ground state of $^{39}$K (Fig.~\ref{fig:apparatus}(a)). This laser serves as a monitor of the production of potassium atoms inside the CBGB source, as well as a frequency reference for the second laser. The second TA Pro system is locked to the first via an offset lock to enable fast and continuous frequency tuning with respect to the crossover frequency. This laser is capable of 2~W of output power and is referred to as the MOT laser. It is split into two components and frequency shifted with acousto-optic modulators (AOMs) to address the cycling $\left(^2S_{1/2}~F'=2 \rightarrow ^2P_{3/2}~F=3\right)$ and the repumping $\left(^2S_{1/2}~F'=1 \rightarrow ^2P_{3/2}~F=2\right)$ transitions. The detuning, $\delta$, of the MOT beams is set by the variable offset lock frequency and thus is the same for both cycling and repumping beams. For absorption imaging of the MOT, we pick off a small amount of light and frequency shift it with an AOM to be resonant with the repumping transition.

 The cycling and repumping light are combined using half-wave plates on polarizing beam splitters and separated into three parts for the MOT axes. Due to the fact that the cycling and repumping light are of orthogonal linear polarizations, it is not possible to combine them in equal ratios among all three MOT beams while maximizing total power at the MOT. Equal power, $\sim$100~mW, is sent through an optical fiber to the vertical arm and combined horizontal arms of the MOT beams. The horizontal arm is then split equally into two orthogonal directions, each at 45$^\circ$ relative to the atomic beam axis, with $\sim$50~mW each. The ratio of repumping to cycling laser power is 1:6 in the vertical arm and 1:4 in each horizontal arm. 
The MOT beams are expanded to a $1/e^2$ diameter of 27~mm 
but limited by apertures to within a 17~mm diameter, presenting a reasonably uniform intensity profile in the MOT region. 
The saturation intensity, $I_{\rm{sat}}$, for the cycling transition of the D$_2$ line in $^{39}$K is 1.75~mW/cm$^2$, so the peak intensity of cycling light is approximately 8$\,I_{\rm{sat}}$ in the horizontal beams and 17$\,I_{\rm{sat}}$ in the vertical beam.

The MOT magnetic field gradient is generated by a set of coils in an anti-Helmholtz configuration. Each coil consists of 2 layers of 10 concentric turns. The inner radius of the coils is 4.2~cm and the distance between the coils is 17.6~cm. At a current of 50~A, we produce a gradient of 4~G/cm in the radial direction and 8~G/cm in the axial direction.  

We measure the forward velocity of the beam in the MOT region by employing one of the horizontal MOT beams at 45$^\text{o}$ relative to the atomic beam axis, with its retro-reflection and the other two arms of the MOT beams blocked. Due to the Doppler shift, the fluorescence observed on a photomultiplier tube (PMT) as a function of detuning, $\delta$, can be mapped to the population of the atomic beam with forward velocity $v$ according to $v=\sqrt{2}c\delta/f_0$, where $c$ is the speed of light and $f_0$ is the resonant transition frequency for an atom at rest. The velocity distribution inferred in this way, $g_0(v)$, is corrected to account for the fact that slower atoms scatter more photons as they traverse the finite region of the laser beam, so that the true velocity distribution is $g(v)\propto g_0(v)\times v$.  The peak forward velocity of the beam is obtained from a Gaussian fit and is 182(1)~m/s (Fig.~\ref{fig:beamtraces}(a)), a typical value for a single-stage atomic CBGB \cite{hutzler2012buffer}. The forward velocity distribution has a full width at half maximum (FWHM) of \blue{90(4)}~m/s. In the regime of relatively low flow rates used here, the spreads of transverse and forward velocities are approximately equal~\cite{hutzler2012buffer}.

To better understand the atomic beam at the MOT location, we measure the absorption of a laser resonant with the cycling transition of the D$_2$ line. The laser passes through the atomic beam 13 times before being recorded on a photodiode, resulting in a peak absorption of approximately \blue{3.6}\% of the laser light. Monte Carlo simulations of the atomic beam propagation are used to determine the transverse density profile of the atoms at the MOT location, allowing us to interpret this result as a peak on-axis density of approximately \blue{$1.2\times10^6$}~atoms/cm$^3$ at the MOT location, 65~cm away from the buffer gas cell. Monte Carlo simulations of the atomic beam additionally show that this peak density of ballistic atoms at the MOT location is consistent with a total production of \blue{$5\times10^{10}$}~atoms/pulse and an on-axis beam brightness of \blue{$2\times10^{11}$}~atoms/sr/pulse.

To see the modest requirements of the Zeeman slower, we consider a simple illustrative model for the capture velocity, $v_c$, of a MOT. We suppose that atoms entering the MOT region can scatter photons over a length $D$, corresponding to the size of the MOT beams. If the velocity is sufficiently slow, $v<v_c$, then it can be slowed to rest before it exits the area of the MOT beams. In an idealized version of the situation in our experiment, two MOT beams propagate at an angle of $\theta=45^\circ$ against the atomic beam, and two MOT beams propagate vertically, perpendicular to the atomic beam. We suppose the retro-reflected horizontal beams propagating at 45$^\circ$ along the atomic beam direction can be neglected since their red-detuning combines with the Doppler effect to shift them far off resonance. The saturated total scattering rate is $\Gamma/2$, where $\Gamma = 2\pi\times 6.035$~MHz is the  natural linewidth of the D$_2$ transition. Only half of the absorbed photons are due to the horizontal, partially counter-propagating beams, and thus the net force against the atomic beam direction is $\hbar k \cos\theta\Gamma/4$, where $\hbar k$ is the momentum that would be transferred by an exactly counter-propagating photon with wavenumber $k$. By kinematics, the maximum initial velocity that can be stopped in distance $D$ is then
\begin{equation}
v_c \approx \sqrt{\frac{\hbar k\Gamma D\cos{\theta}}{2m}}.
\end{equation}

The two horizontal MOT beam arms, each \blue{17}~mm in diameter, overlap in a region of length $D=\blue{25}$~mm along the atomic beam axis but with $\theta=\pm45^\circ$. The estimated capture velocity is thus \blue{67}~m/s.
From the velocity distribution of the atomic beam, we see that a negligible fraction of atoms are produced with velocities below $v_c$, but a reduction in velocity of only $\sim$100~m/s is sufficient to bring a significant fraction of produced atoms to within the capture range of the MOT. The slower we use is schematically shown in Fig. \ref{fig:apparatus}(b). It is 11~cm long and consists of three concentric layers of length 11~cm, 9~cm and 5~cm. The slower is operated at a current of 10~A, where it produces a maximum magnetic field of 110~G. Because of the relatively low  DC currents used to drive both the MOT and Zeeman slower coils, no water cooling is required to prevent overheating. To generate the light for slowing, we use a Titanium:Sapphire ring laser. We add a sideband with an AOM to address the repumping transition and send 180~mW of collimated slowing light (1:2 repumping to cycling power ratio) counter-propagating along the atomic beam axis.

Employing the Zeeman slower, we are able to load a MOT and detect it with a PMT as shown in Fig.\ref{fig:beamtraces}(b). 
The optimal detuning for the MOT is -14~MHz (-2.3~$\Gamma$) and the optimal detuning for the Zeeman slower is -59~MHz (-9.8~$\Gamma$). The use of a PMT allows us to obtain time-resolved information on the loading rate and the decay of the MOT. We note that the peak of the MOT signal appears around 12~ms after the YAG pulse. The time trace of an unperturbed beam of potassium atoms, observed via laser-induced fluorescence from the MOT lasers, is also depicted. 

In the case of a pulse-loaded MOT, the density $n(t)$ at time $t$ can be obtained by solving the differential equation \cite{Hemmerling2014},
\begin{equation}\label{eq:loading-model}
    \frac{dn}{dt} = R(t) - \alpha n(t) - \beta n(t)^2,
\end{equation}
where $R(t)$ is the loading rate from a pulsed source, $\alpha$ is the background gas collisional loss rate, and $\beta$ is the two-body collisional loss rate. Since the densities for which two-body loss is non-negligible are much higher than what we directly load in our MOT \cite{Weiner1999}, we may omit $\beta$ in the following treatment. The solution of the above differential equation is then
\begin{eqnarray}\label{eq:profile}
    n(t) &=& \frac{1}{2} n_{\rm{tot}}e^{-\alpha(t - t_0 - \alpha w^2/2)}\times\\ \nonumber
    &&\left[\text{erf}\left(\frac{t_0+\alpha w^2}{\sqrt{2}w}\right)-\text{erf}\left(\frac{-t + t_0 + \alpha w^2}{\sqrt{2}w}\right)\right],
\end{eqnarray}
where we have assumed the atomic pulse loading rate $R(t)$ to be a Gaussian function centered at $t=t_0$ with standard deviation $w$. The fit for the MOT in Fig.~\ref{fig:beamtraces}(b) is overlaid to the data and shows good agreement with the measurement. At times $t\gtrsim t_0+2w$, the density approaches the form $n(t)\propto \exp{(-\alpha t)},$ such that the lifetime is $\tau=1/\alpha$. Extracting the MOT lifetime from the full fit above, we obtain a lifetime of $\tau=$116.4(4)~ms. The characteristic loading time $w$ is found to be \blue{2.6}~ms, so that 95\% of atoms are loaded into the MOT within a \blue{10.3}~ms window.

\begin{figure}[ht]
\begin{centering}
\includegraphics[width = 1\columnwidth]{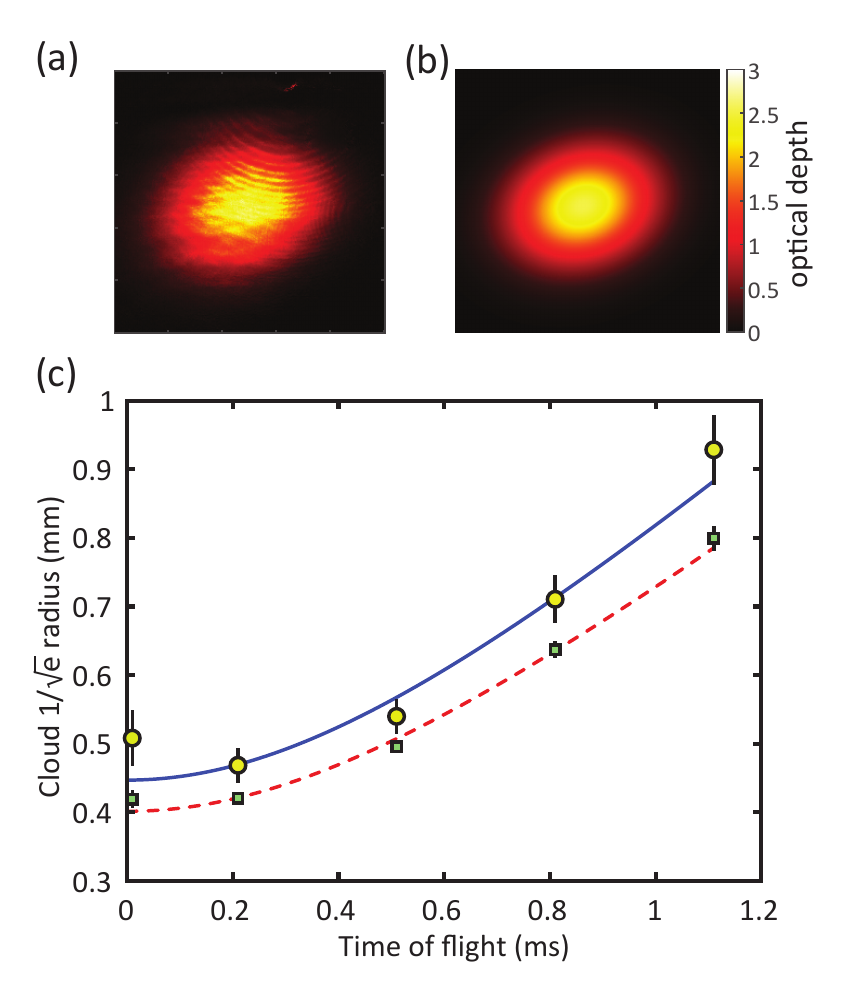}
\caption{\label{fig:motDensityAndT} \textbf{MOT absorption images and temperature}. (a) Optical density from absorption images of the MOT after \blue{910}~$\mu$s time of flight and averaged over \blue{36} shots. (b) Fit of absorption image to a two-dimensional Gaussian function. (c) Time-of-flight trace of the $1\sigma$ radius obtained from the Gaussian fits in each direction (horizontal in yellow circles, vertical in green squares), with best-fit curves shown for the model $\sigma^2(t) = \sigma_0^2 + k_BTt^2/m$.}
\par\end{centering}
\end{figure}

To extract the MOT density and temperature, we perform absorption imaging along the horizontal direction orthogonal to the atomic beam axis. We use an imaging system of 3$\times$ demagnification and a CCD camera. We turn off the MOT magnetic field after the entire pulse has been loaded into the MOT, typically 15~ms after the YAG pulse. 
The time-of-flight is defined starting from this instant. We send $\sim$5~$\mu$W of light resonant with the repumping transition to obtain absorption images like the ones shown in Fig.~\ref{fig:motDensityAndT}(a). The imaging light is polarized along the quantization axis defined by the weak fringing field of the Zeeman slower 14 cm away, which for the repumping transition leads to a resonant optical scattering cross-section of \blue{$\sigma=7.0\times10^{-10}$~cm$^2$}.
From the integrated optical depth and cross-section, we obtain the total number of atoms. 

We fit the optical depth, $OD$, of the absorption images to a two-dimensional Gaussian distribution,
\begin{equation}
    OD=OD_0\exp{\left(-\frac{(x-x_0)^2}{2\sigma_x^2}-\frac{(y-y_0)^2}{2\sigma_y^2}\right)},
\end{equation}
to obtain the size of the MOT defined as the $1\sigma$ (equivalently, $1/\sqrt{e}$) radius in the two directions (Fig.~\ref{fig:motDensityAndT}(b)). We repeat this process for expansion times up to 1.2~ms. The size of the MOT as a function of time-of-flight is shown in Fig.~\ref{fig:motDensityAndT}(c). We fit the points to the expression $\sigma^2(t) = \sigma_0^2 + k_BTt^2/m$ to obtain the MOT temperature $T$ and the MOT in-trap size $\sigma_0$. We obtain a temperature of 2.2(3)~mK in the horizontal direction and 1.7(1)~mK in the vertical. The initial sizes, $\sigma_0$, are \blue{0.447(22)}~mm in the horizontal direction and \blue{0.401(7)}~mm in the vertical direction. The MOT temperatures in orthogonal directions are consistent within measurement uncertainty, and are in good agreement with expectations for the relatively large values of $I/I_{\rm{sat}}$ used here~\cite{Phillips1990}. We report trapped atom numbers from measurements with approximately 1~ms time-of-flight expansion, for which saturation effects in the absorption images are negligible and the integrated optical depth has lower uncertainty. We measure atom numbers as high as \blue{$1.05(1)\times10^8$}, corresponding to the in-trap density \blue{$8.3(8)\times10^{10}$}~atoms/cm$^3$. This corresponds to a phase space density of \blue{$2\times10^{-8}$}. Combining the peak atom number in the MOT with PMT traces (Fig.~\ref{fig:beamtraces}(b)) fit to the model of Eq.~\ref{eq:loading-model}, we infer a peak loading rate of \blue{$1.8\times10^{10}$}~atoms/s.

\begin{figure}[ht]
\begin{centering}
\includegraphics[width = 1\columnwidth]{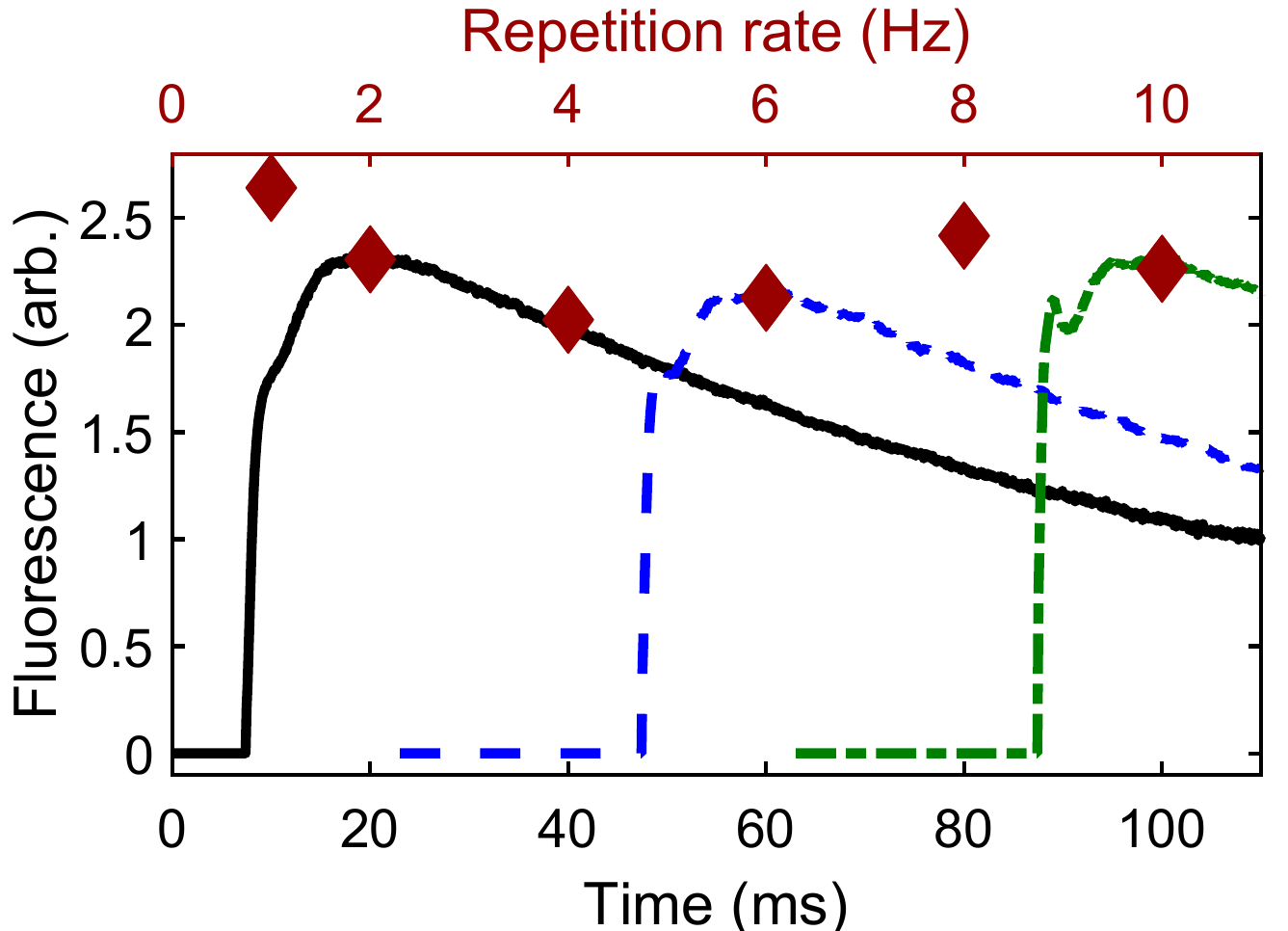}
\caption{\label{fig:rep-rate} \textbf{MOT fluorescence as a function of repetition rate.} For clarity, PMT traces are shown offset in time only for 2 (black, solid), 6 (dashed, blue), and 10 (dot-dashed, green) Hz repetition rates. Peak fluorescence signals (red diamonds), proportional to peak MOT densities, are indicated for 1, 2, 4, 6, 8, and 10 Hz. Fluctuations of $\approx$10\% in peak signal are typical of ablation yields in CBGB sources. At the fastest repetition rates, due to the short experimental cycle time (at minimum, 100 ms), the tail of the MOT decay curve is not recorded. Transient spikes near the MOT turn-on time are due to scattered light arising from switching lasers; signal before these transients is set to 0 for readability.
}
\par\end{centering}
\end{figure}

Finally, we measure the MOT loading efficiency and lifetime as a function of the repetition rate (Fig.~\ref{fig:rep-rate}). We observe no strong dependence of the peak MOT density on the experimental repetition rate between 1 and 10~Hz. We measure a <10\% standard deviation among the peak densities observed with a PMT, consistent with the typical scale of fluctuations in ablation yield. 

We can estimate the expected MOT lifetime by modeling the effect of collisions between potassium and helium atoms, assuming a trapped atom is knocked out of the MOT by a single collision with a background helium atom at room temperature. Thus the expected decay rate is given by $\alpha = n_{\rm{He}}\,\bar{v}_{\rm{He}}\,\sigma_{\rm{He-K}}$, where $n_{\rm{He}}$ is the ambient helium density in the MOT region, $\bar{v}_{\rm{He}}=\sqrt{8k_{\rm{B}}T/\pi\,m_{\rm{He
}}}$ is the average velocity of a helium atom at room temperature, and $\sigma_{\rm{He-K}}=1.7\times10^{-14}$~cm$^2$ is the helium-potassium collision cross-section~\cite{Rosenberg1939}. We estimate $n_{\rm{He}}$ to be around $3\times10^9$~atoms/cm$^3$ from the equilibrium pressure of approximately $1\times10^{-7}$~Torr. We obtain an estimated value of $1/\alpha=150$~ms, in reasonable agreement with the measured lifetimes. We have additionally verified a linear dependence of the trapped atom loss rate with helium background pressure over more than an order of magnitude.

Although the MOT density after direct loading is quite high, it can be further increased with standard techniques. The first method is compression wherein the MOT laser intensity is ramped down in $\sim$10~ms simultaneously with an increasing ramp of the magnetic field gradient \cite{Petrich1994}. This number-conserving technique cools the MOT closer to the Doppler limit. For our demonstrated trapped atom numbers, phase-space densities greater than $10^{-7}$ can be expected in a compressed MOT near the Doppler limit, and even higher phase-space densities are achievable with sub-Doppler techniques~\cite{Salomon2013}. 
A complementary strategy to increase the loaded atom number and density is to collimate the atomic beam in the cryogenic buffer gas beam source chamber using techniques like transverse Doppler cooling or a two-dimensional MOT \cite{Dieckmann1998}. Since the atomic beam expands as it traverses the 65~cm between the cell and the MOT, the fraction of the cell yield that reaches the MOT depends on the solid angle subtended. By collimating the atomic beam, one can increase the number within the MOT capture volume by a factor of \blue{90}, 
estimated with Monte Carlo particle trajectory simulations. This would require installing mirrors and magnetic field coils within the buffer gas beam source vacuum chamber. The implementation of these techniques is beyond the scope of this work but can be considered as simple upgrades to the system in order to achieve trapped atom numbers well above $10^{9}$ and densities higher than $10^{13}$~atoms/cm$^3$ per pulse. We emphasize that these techniques remain compatible with the \blue{10}~ms MOT loading time and 10~Hz experimental repetition rate demonstrated here.

In conclusion, we have demonstrated pulsed MOT loading of $^{39}$K atoms from a cryogenic buffer gas beam source. We load a MOT of $10^8$~atoms in \blue{10}~ms. The densities routinely obtained are on the order of $\blue{10^{11}}$~atoms/cm$^3$. The system operates at repetition rates as high as 10~Hz without measurable loss in MOT density. The MOT lifetime is approximately 120~ms, which is sufficient for sub-Doppler cooling and trapping in tweezers. We establish that this lifetime is limited purely by buffer gas leakage out of the CBGB and thus can be mitigated by adding a differential pumping stage between the CBGB and the MOT region if needed. The achieved densities are on par with the state-of-the-art for D$_2$ MOTs operating with traditional oven or dispenser sources, but with a duty cycle more than an order of magnitude higher and one hundred times shorter loading time. We note that the measured densities are obtained without compression, and could be improved by cooling the MOT cloud to the Doppler limit using standard compression techniques. Furthermore, Monte Carlo simulations show that a two-dimensional MOT in the CBGB chamber would additionally increase the flux of atoms below the capture velocity at the MOT by a factor of \blue{$\sim$100}, which would enable a comparable number of loaded atoms to the highest-yield oven-based sources using similar methods~\cite{Ridinger2011}, but in a much shorter loading time. The high densities and duty cycles achieved with this approach make it ideally suited for experiments where large data rates are required. Finally, a CBGB source can be used to load a MOT of almost any species and the atomic flux is not limited by vapor pressure, making the methods demonstrated here especially advantageous for quickly loading high-density traps of refractory metals such as chromium and titanium.

This work was supported by the Harvard University Dean's Competitive Fund for Promising Scholarship. We would like to acknowledge L. Anderegg and Y. Bao for insightful discussions.

\bibliographystyle{apsrev4-2}
\bibliography{KMOT_References}
\end{document}